\documentclass[12pt]{article}
\usepackage{graphics,psfrag}
\usepackage{amsthm,amssymb,epsfig,amsmath,euscript,array,cite}

\newcounter{multieqs}

\newcommand{\be}{\begin{equation}}
\newcommand{\ee}{\end{equation}}
\newcommand{\eq}[1]{(\ref{#1})}

\newcommand{\bm}[1]{\mbox{\boldmath $#1$}}
\newcommand{\rf}[1]{(\ref{#1})}

\def\bd{\begin{document}}
\def\ed{\end{document}}
\def\nn{\nonumber}
\def\bea{\begin{eqnarray}}
\def\eea{\end{eqnarray}}
\let\bm=\bibitem
\let\la=\label

\def\npb#1#2#3{Nucl. Phys. {\bf{B#1}} #3 (#2)}
\def\plb#1#2#3{Phys. Lett. {\bf{#1B}} #3 (#2)}
\def\prl#1#2#3{Phys. Rev. Lett. {\bf{#1}} #3 (#2)}
\def\prd#1#2#3{Phys. Rev. {D \bf{#1}} #3 (#2)}
\def\cmp#1#2#3{Comm. Math. Phys. {\bf{#1}} #3 (#2)}
\def\cqg#1#2#3{Class. Quantum Grav. {\bf{#1}} #3 (#2)}
\def\nppsa#1#2#3{Nucl. Phys. B (Proc. Suppl.) {\bf{#1A}}#3 (#2)}
\def\ap#1#2#3{Ann. of Phys. {\bf{#1}} #3 (#2)}
\def\ijmp#1#2#3{Int. J. Mod. Phys. {\bf{A#1}} #3 (#2)}
\def\rmp#1#2#3{Rev. Mod. Phys. {\bf{#1}} #3 (#2)}
\def\mpla#1#2#3{Mod. Phys. Lett. {\bf A#1} #3 (#2)}
\def\jhep#1#2#3{J. High Energy Phys. {\bf #1} #3 (#2)}
\def\atmp#1#2#3{Adv. Theor. Math. Phys. {\bf #1} #3 (#2)}

\newcommand{\EQ}[1]{\begin{equation} #1 \end{equation}}
\newcommand{\AL}[1]{\begin{subequations}\begin{align} #1 \end{align}
\end{subequations}}
\newcommand{\SP}[1]{\begin{equation}\begin{split} #1 \end{split}\end{equation}}
\newcommand{\ALAT}[2]{\begin{subequations}\begin{alignat}{#1} #2 
\end{alignat}\end{subequations}}
\def\beqa{\begin{eqnarray}} 
\def\eeqa{\end{eqnarray}} 
\def\beq{\begin{equation}} 
\def\eeq{\end{equation}} 

\def\N{{\cal N}}
\def\sst{\scriptscriptstyle}
\def\thetabar{\bar\theta}
\def\Tr{{\rm Tr}}
\def\one{\mbox{1 \kern-.59em {\rm l}}}

\def\a{\alpha}      \def\da{{\dot\alpha}}  
\def\b{\beta}       \def\db{{\dot\beta}}  
\def\c{\gamma}  \def\C{\Gamma}  \def\dc{{\dot\gamma}}  
\def\d{\delta}  \def\D{\Delta}  \def\ddt{\dot\delta}  
\def\e{\epsilon}        \def\vare{\varepsilon}  
\def\f{\phi}    \def\F{\Phi}    \def\vvf{\f}  
\def\h{\eta}  
\def\k{\kappa}  
\def\l{\lambda} \def\L{\Lambda}  
\def\m{\mu} \def\n{\nu}  
\def\o{\omega}  
\def\p{\pi} \def\P{\Pi}  
\def\r{\rho}  
\def\s{\sigma}  \def\S{\Sigma}  
\def\t{\tau}  
\def\th{\theta} \def\Th{\Theta} \def\vth{\vartheta}  
\def\X{\Xeta}  
\def\z{\zeta}  
  
\def\cA{{\cal A}} \def\cB{{\cal B}} \def\cC{{\cal C}}  
\def\cD{{\cal D}} \def\cE{{\cal E}} \def\cF{{\cal F}}  
\def\cG{{\cal G}} \def\cH{{\cal H}} \def\cI{{\cal I}}  
\def\cJ{{\cal J}} \def\cK{{\cal K}} \def\cL{{\cal L}}  
\def\cM{{\cal M}} \def\cN{{\cal N}} \def\cO{{\cal O}}  
\def\cP{{\cal P}} \def\cQ{{\cal Q}} \def\cR{{\cal R}}  
\def\cS{{\cal S}} \def\cT{{\cal T}} \def\cU{{\cal U}}  
\def\cV{{\cal V}} \def\cW{{\cal W}} \def\cX{{\cal X}}  
\def\cY{{\cal Y}} \def\cZ{{\cal Z}}

\def\ua{\underline{\alpha}}  
\def\ub{\underline{\phantom{\alpha}}\!\!\!\beta}  
\def\uc{\underline{\phantom{\alpha}}\!\!\!\gamma}  
\def\um{\underline{\mu}}  
\def\ud{\underline\delta}  
\def\ue{\underline\epsilon}  
\def\una{\underline a}\def\unA{\underline A}  
\def\unb{\underline b}\def\unB{\underline B}  
\def\unc{\underline c}\def\unC{\underline C}  
\def\und{\underline d}\def\unD{\underline D}  
\def\une{\underline e}\def\unE{\underline E}  
\def\unf{\underline{\phantom{e}}\!\!\!\! f}\def\unF{\underline F}  
\def\unm{\underline m}\def\unM{\underline M}  
\def\unn{\underline n}\def\unN{\underline N}  
\def\unp{\underline{\phantom{a}}\!\!\! p}\def\unP{\underline P}  
\def\unq{\underline{\phantom{a}}\!\!\! q}  
\def\unQ{\underline{\phantom{A}}\!\!\!\! Q}  
\def\unH{\underline{H}}  
  
\def\As {{A \hspace{-6.4pt} \slash}\;}  
\def\bs {{b \hspace{-6.4pt} \slash}\;}  
\def\Ds {{D \hspace{-6.4pt} \slash}\;}  
\def\ds {{\del \hspace{-6.4pt} \slash}\;}  
\def\ss {{\s \hspace{-6.4pt} \slash}\;}  
\def\ks {{ k \hspace{-6.4pt} \slash}\;}  
\def\ps {{p \hspace{-6.4pt} \slash}\;}  
\def\pas {{{p_1} \hspace{-6.4pt} \slash}\;}  
\def\pbs {{{p_2} \hspace{-6.4pt} \slash}\;}  
  
\def\Fh{\hat{F}}  
\def\Vh{\hat{V}}  
\def\Xh{\hat{X}}  
\def\ah{\hat{a}}  
\def\xh{\hat{x}}  
\def\yh{\hat{y}}  
\def\ph{\hat{p}}  
\def\xih{\hat{\xi}}  
  
\def\psit{\tilde{\psi}}  
\def\Psit{\tilde{\Psi}}   
\def\Psibt{\tilde{\bar{Psi}}}  

\def\Phit{\tilde{\Phi}}   
\def\Phitb{\overline{\tilde{Phi}}}  
\def\tht{\tilde{\th}}  
   
\def\At{\tilde{A}}  
\def\Qt{\tilde{Q}}  
\def\Rt{\tilde{R}}  
\def\Nt{\tilde{N}}  
 
\def\at{\tilde{a}}  
\def\st{\tilde{s}}  
\def\ft{\tilde{f}}  
\def\pt{\tilde{p}}  
\def\qt{\tilde{q}}  
\def\vt{\tilde{v}}  
\def\nt{\tilde{n}}  
  
\def\delb{\overline{\partial}}  
\def\thb{\overline{\theta}}
\def\mub{{\overline \mu}}
\def\lamb{{\overline \l}}
\def\psib{{\overline \psi}}
\def\sb{{\overline \sigma}}
\def\xib{{\overline \xi}}
\def\chib{{\overline \chi}}

\def\Phib{\overline{\Phi}}
\def\Lamb{\overline{\Lambda}}

\def\Ab{{\overline A}} \def\Bb{{\overline B}} \def\Cb{{\overline C}}  
\def\Db{{\overline D}} \def\Eb{{\overline E}} \def\Fb{{\overline F}}  
\def\Gb{{\overline G}} \def\Hb{{\overline H}} \def\Ib{{\overline I}}  
\def\Jb{{\overline J}} \def\Kb{{\overline K}} \def\Lb{{\overline L}}  
\def\Mb{{\overline M}} \def\Nb{{\overline N}} \def\Ob{{\overline O}}  
\def\Pb{{\overline P}} \def\Qb{{\overline Q}} \def\Rb{{\overline R}}  
\def\Sb{{\overline S}} \def\Tb{{\overline T}} \def\Ub{{\overline U}}  
\def\Vb{{\overline V}} \def\Wb{{\overline W}} \def\Xb{{\overline X}}  
\def\Yb{{\overline Y}} \def\Zb{{\overline Z}}  

\def\fb{{\overline f}}
\def\gb{{\overline g}}
\def\mb{{\overline m}}
\def\lb{{\overline l}}
\def\yb{{\overline y}}

\def\ba{{\bf a}} 
\def\bk{{\bf k}}  
\def\bl{{\bf l}}  
\def\bp{{\bf p}}  
\def\bq{{\bf q}}  
\def\br{{\bf r}}
\def\bt{{\bf t}}
\def\bu{{\bf u}}
\def\bv{{\bf v}}
\def\bx{{\bf x}}  
\def\by{{\bf y}}  
\def\bR{{\bf R}}  
\def\bV{{\bf V}}

\def\bone{{\bf 1}}  

\def\va{{\vec a}}
\def\vp{{\vec p}}
\def\vq{{\vec q}}
\def\vx{{\vec x}}
\def\vu{{\vec u}}
\def\vv{{\vec v}}

\def\vs{{\vec \sigma}}
\def\vtau{{\vec \tau}}

\newcommand{\ov}[1]{\overrightarrow{#1}}
  
\def\d{\delta}\def\D{\Delta}\def\ddt{\dot\delta}  
  
\def\pa{\partial} \def\del{\partial}  
\def\xx{\times}  
\def\uno{\mbox{1 \kern-.59em {\rm l}}}    
  
\def\trp{^{\top}}  
\def\inv{^{-1}}  
\def\dag{{^{\dagger}}}  
\def\pr{^{\prime}}  
  
\def\rar{\rightarrow}  
\def\lar{\leftarrow}  
\def\lrar{\leftrightarrow}  
  
\newcommand{\0}{\,\!}      
\def\one{1\!\!1\,\,}  
\def\im{\imath}  
\def\jm{\jmath}  
  
\newcommand{\tr}{\mbox{tr}}  
\newcommand{\slsh}[1]{/ \!\!\!\! #1}  
  
\def\vac{|0\rangle}  
\def\lvac{\langle 0|}  
  
\def\hlf{\frac{1}{2}}  
\def\ove#1{\frac{1}{#1}}  

\def\Box{\square}  
\def\ZZ{\mathbb{Z}}  
\def\bb#1{{\bf #1}}  
\def\bcomment#1{}  
\def\bfhat#1{{\bf \hat{#1}}}  
\def\VEV#1{\left\langle #1\right\rangle}  

\newcommand{\ex}[1]{{\rm e}^{#1}} \def\ii{{\rm i}}  

\newcommand{\lrbrk}[1]{\left(#1\right)}
\newcommand{\sfrac}[2]{{\textstyle\frac{#1}{#2}}}
 
\def\stw{{\sqrt{2}}}

\def\rf {{\rm f}}
\def\ri {{\rm i}}
\def\rs {{\scriptscriptstyle \rm S}}
\def\rt {{\scriptscriptstyle \rm T}}

\def\rQ {{\scriptscriptstyle \rm \cQ}}
\def\rR {{\scriptscriptstyle \rm \cR}}

\def\cQb{{\cal \Qb}}
\def\cRb{{\cal \Rb}}
\def\cWb{{\cal \Wb}}

\def\fd {{\rm N}}
\def\afd {{\overline{\rm N}}}

\font\myBB=msbm10 at 18pt
\def\BB#1{\hbox{\myBB#1}}

\begin{document}

\hfill{hep-th/0505141}

\vspace{20pt}

\begin{center}

{\Large \bf  Konishi Anomaly and Central Extension 
in \\ 
$\cN =\frac{1}{2}$ Supersymmetry} 
\vspace{30pt}

{\bf Chong-Sun Chu$^{a}$, Takeo Inami$^{b}$}

\vspace{15pt}
{\small \em
\begin{itemize}
\item[$^a$]
Centre for Particle Theory and Department of Mathematics,
University of Durham, Durham, DH1 3LE, UK.
\item[$^b$]
Department of Physics, Chuo University, Kasuga, Bunkyo-ku, Tokyo, 
112-8551, Japan.
\end{itemize}
}

\vskip .1in {\small \sffamily chong-sun.chu@durham.ac.uk, 
inami@phys.chuo-u.ac.jp}

\vspace{50pt}
{\bf Abstract}

\end{center}

We show that the 4-dimensional $\cN=1/2$ supersymmetry algebra admits
central extension. The central charges are supported by domain wall
and the central charges are computed. We also determine the Konishi
anomaly for $\cN=1/2$ supersymmetric gauge theory. Due to the new
couplings in the Lagrangian, many terms appears. We show that 
these terms sum up to give the expected form for the
holomorphic part of the Konishi anomaly. For the anti-holomorphic part,
we give a simple argument that the naive generalization has to be
modified. We suggest that the anti-holomorphic Konishi anomaly is given
by a gauge invariant completion using open Wilson line. 

\setcounter{page}0
\newpage

\section{Introduction}

Quantum field theory on noncommutative space, $[x^\m,x^\n] = i \th^{\m\n}$,
displays a rich spectrum of
unusual properties,  some of which are believed to be relevant for
quantum gravity \cite{review}. 
A natural extension of the noncommutative space is to
consider deformed superspace. Superspace with only the
bosonic coordinates deformed were considered in
\cite{susy}. The more general superspace where
the Grassmann odd part is deformed was also  
considered in \cite {more-susy}.
More recently, starting with the observation of Ooguri and Vafa \cite{ov},
string theory in graviphoton background has been considered 
and it is found that a self-dual graviphoton field strength $C_{\m\n}$
induces a deformation of the 4-dimensional superspace 
\cite{ov,bgn,seiberg,bs} so
that the Grassmann odd coordinates become non-anticommutative. 
In particular, the deformation keeps $\cN=1/2$ supersymmetry
\cite{seiberg,bs}.  

Supersymmetric quantum field theory on
non-anticommutative superspace
was first formulated by Seiberg \cite{seiberg} and 
deformed $\cN=1/2$ Wess-Zumino model and pure SYM were constructed.
Various generalizations are possible. 
$\cN=1/2$ supersymmetric gauge theory with chiral matters was
constructed in \cite{ito}, where the modification to the supersymmetry
transformations of the chiral matter fields were determined. See also
\cite{further} for further studies.
Nonlinear sigma models (in four or
two-dimensions) were considered in \cite{nlsm1} and it was found
that the non-anticommutative deformation induces in the Lagrangian 
an infinite number of terms in powers of the auxiliary
field. It turns out that the infinite series can be 
summed up \cite{abbp,agvm,ch} and quite remarkably it can be 
written  in terms of a simple
smearing of the Zumino's Lagrangian and the holomorphic 
superpotential \cite{abbp,agvm}. 
Also, gauge theories with extended supersymmetry
have been constructed using deformed  harmonic superspace
\cite{susy2}, and \cite{susy4} for the deformed $\cN =4$ SYM. 
Instantons have been studied \cite{instanton,imaa}.

The above studies are classical. The quantum properties of non-anticommutative
supersymmetric theories are interesting and important. 
Non-anticommutative supersymmetric theories
are defined in Euclidean  space and are non-hermitian. A priori these
theories can have quite different quantum properties from their
undeformed cousins due to their different structure in supersymmetry.
For the simple $\cN=1/2$ case, it
has been argued that  the Wess-Zumino model and the supersymmetric
gauge theory are renormalizable  in the sense that only a finite
number of counterterms is needed to be added to the original Lagrangian
\cite{renorm1,renorm2}. Some non-renormalization theorems have been 
argued to remain valid. 
Also the one loop effective potential of the Wess-Zumino model 
has been constructed  \cite{wz1loop}. 
Further studies of quantum properties of non-anticommutative theories 
beyond these aspects of renormalizability are however in order. 

At the level of supersymmetry algebra, non-anticommutativity modifies
the anticommutator of  the $\Qb$'s, see \eq{qbar-qbar} below. 
In standard undeformed supersymmetry, it is well
known that at least $\cN=2$ supersymmetry is needed in order to admit
a central extension \cite{haag,wo}. However this holds true when one assumes 
Lorentz
symmetry is unbroken. With Lorentz symmetry broken, one can actually
have a central extension in the undeformed $\cN=1$ supersymmetry
algebra. The central charge is carried by a domain wall \cite{DS,CS}. For
special configuration, the wall is BPS and half of the supersymmetries
are left unbroken. It is interesting to see whether the deformed 
$\cN=1/2$ supersymmetry algebra also admits central extension, and
how it is affected by the non-anticommutativity. 
In section 3, we show that central extension is possible in the
$\cN=1/2$ Wess-Zumino model. We construct the domain wall and show
that it breaks all the supersymmetries. We also show that the central
charge is unaffected by the presence of $C^{\a\b}$.

Another purpose of this paper is to 
study the quantum properties of 4-dimensional non-anticommutative
supersymmetric theories.  In section 4, we carry out a detailed analysis
of the Konishi
anomaly in $\cN=1/2$ supersymmetric gauge theory. This anomaly arises
in one-loop. We will find that the holomorphic Konishi anomaly  takes
the expected form (i.e. dressing up the usual relation with
$*$-product), while the  anti-holomorphic Konishi
anomaly is nontrivially modified. In the next section, we will begin
with  a brief review of the properties of the $\cN=1/2$ superspace. 
Discussion of our results and further directions of investigation are
given in section 5.

\section{$\cN =1/2$ Superspace}

Let $(x^\m, \th^\a, \thb^\da)$ be the coordinates of the
4-dimensional non-anticommutative superspace \cite{seiberg}. When  a
graviphoton background is turned on, the superspace coordinates  obey
the relations 
\bea 
& \{\thb^\da, \thb^\db \} = \{\thb^\da, \th^\b \} =0, 
\quad \{ \th^\a, \th^\b\} = C^{\a\b}, \\ 
& [y^\m,y^\n] = [y^\m,\th^\a] = [y^\m, \thb^\da] =0, 
\eea 
where  $y^\m = x^\m + i \th \s^\m \thb$ 
is the chiral coordinate.  Functions of $\th$ are Weyl ordered
using the $*$-product 
\be 
f(\th) * g(\th) = f(\th) \exp\left(
-\frac{C^{\a\b}}{2} \overleftarrow{\frac{\del}{\del\th^\a}}
\overrightarrow{\frac{\del}{\del\th^\b}} \right) g(\th).  
\ee
As is obvious from the above relations, $\thb$ is not the
complex conjugate of $\th$. The deformation is possible
only for Euclidean space or $(2,2)$-signature.
We will be working in Euclidean space and we  follow the convention of
\cite{seiberg} to continue to use the Lorentzian signature notation.
The  $(2,2)$-signature is relevant for $\cN=2$ string theory and 
for the studies of non-anticommutative version of
supersymmetric integrable systems.

Written in the chiral basis $y, \th, \thb$,  the supercharges and
covariant derivatives take the standard expressions 
\bea 
&&Q_\a = \frac{\del}{\del \th^\a}, \qquad \quad  
\Qb_{\da} = -\frac{\del}{\del \thb^\da} 
+ 2 i \th^\a \s^\m_{\a\da} \frac{\del}{\del y^\m},  \\ 
&&D_a = \frac{\del}{\del \th^\a}  + 2 i \s^\m_{\a\da} \thb^{\da}
\frac{\del}{\del y^\m}, \quad  
\Db_{\da} = -\frac{\del}{\del \thb^\da}.
\eea
They satisfy 
\bea 
&& \{ Q_\a, \Qb_\da\} = 2i \s^\m_{\a\da} \frac{\del}{\del y^\m}, \\ 
&& \{ Q_\a, Q_\b\} = 0, \\ 
&& \{ \Qb_\da, \Qb_\db\} = - 4 C^{\a \b} \s^\m_{\a\da} \s^\n_{\b\db}
\frac{\del^2}{\del y^\m \del y^\n}, \label{qbar-qbar}
\eea 
and 
\bea  
&& \{ D_\a, \Db_\da\} = -2i \s^\m_{\a\da} \frac{\del}{\del y^\m}, \\ 
&& \{ D_\a, D_\b\} = \{ \Db_\da, \Db_\db\} =0, 
\eea 
with all the remaining
anti-commutators equal to zero. Due to the dependence of $\Qb$'s on
the non-anticommutative coordinates $\th$, $\Qb$ is no longer a
symmetry of the noncommutative superspace. The  $\cN =1/2$
supersymmetry is generated by the unbroken $Q$'s.

Chiral (resp. anti-chiral) superfields are  defined by $\Db_\da \Phi
=0$ (resp.  $D_\a \Phib =0$)  and are given by the expansion: 
\be \label{chiral-1} 
\Phi(y, \th) = A(y) + \stw \th \psi(y) + \th\th F(y),
\ee 
\be \label{antichiral-1} 
\Phib(\yb, \thb)=  \Ab(\yb) + \stw \thb \psib(\yb) + \thb\thb \Fb(\yb), 
\ee 
where $\yb^\m = y^\m - 2 i \th^\a \s^\m_{\a\da} \thb^\da$. 
In the presence of gauge symmetry, it is more convenient to
parametrize the anti-chiral fields slightly differently, see
\eq{Sbar}, \eq{Tbar} below, so that the
component fields have the standard form of gauge transformation.

\section{Wess-Zumino Model: Central Charge and Domain Wall}

Consider the $\cN =1/2$ Wess-Zumino model, 
\be 
\cL = \int d^4 \th \Phib* \Phi + \int d^2 \th  \cW(\Phi) + \int d^2
\thb \; \cWb(\Phib)
\ee 
with superpotential ($\l, \lamb >0$), 
\bea \label{cubic-W}
\cW(\Phi) = \mu^2 \Phi + \frac{m}{2} \Phi * \Phi - \frac{\l}{3} \Phi
*\Phi * \Phi, \nn\\ 
\cWb(\Phib) = \mub^2 \Phib + \frac{\mb}{2} \Phib* \Phib 
-  \frac{\lamb}{3} \Phib *\Phib * \Phib.  
\eea
Without loss of generality, one can take $m = \mb =0$. We have
\cite{seiberg} 
\be 
\cL = \cL (C=0) + \frac{1}{3} \l \det C \; F^3, 
\ee
where
\footnote{We normalize $\int d^2 \th \,\th^2 = \int d^2 \thb\, \thb^2
=1$.}  
\bea 
\cL (C=0) &=& \del_\m \Ab \del^\mu A + i \psi^\a \del_{\a\da}
\psib^\da 
+ \Fb F + F \cW'(A)  - \frac{1}{2} \cW''(A) \psi \psi \nn\\ 
&&+ \Fb \cWb'(\Ab) - \frac{1}{2} \cWb''(\Ab) \psib
\psib.  
\eea 
The bosonic equations of motion take the form\footnote{
Here $\Box$ is the (minus) Laplacian in the Euclidean space.}
\bea  
& \Box \Ab = F \cW'', \qquad \Box A = \Fb \,\cWb'',\label{eom-wz-1} \\ 
& F +\cWb' =0, \qquad \Fb + \cW' + \l \det C \; (\cWb')^2 =0.
\label{eom-wz-2} 
\eea 
Note that those for $A$, $\Fb$ are modified by $C$, while those
of $\Ab, F$  are left unchanged. The equations of motion for the
fermions are also unmodified by $C$.

The 
transformations  
$ \d \Phi := (\xi Q + \xib \,\Qb)*\Phi$,  
$\d \Phib := (\xi Q + \xib \,\Qb) * \Phib $  
translate to that of the component fields as 
\bea  
&& \d A = \stw \xi \psi +
\stw i \xib_\dc (\sb^\m)^{\dc \c} \e_{\c\a}  C^{\a\b} \del_\m \psi_\b,
\label{susy-transf-wz1} \\ 
&& \d \psi = \stw \xi F + 2 i \s^\m \xib
\del_\m A, \label{susy-transf-wz2} \\ 
&& \d F = \stw i \xib \sb^\m
\del_\m \psi,  \label{susy-transf-wz3} 
\eea 
and 
\bea 
&&\d \Ab = \stw \xib \psib, \label{susy-transf-wz4}\\ 
&&  \d \psib = \stw \xib \Fb + \stw i \sb^\mu \xi \del_\m \Ab,
\label{susy-transf-wz5}\\ 
&& \d \Fb = \stw i \xi \s^\m \del_\m \psib.  \label{susy-transf-wz6} 
\eea 
It is evident
that the theory is invariant under the  $Q$-supersymmetry and 
broken for the $\Qb$-transformation. One can
easily work out the conserved supercurrent
\be \label{J-wz} 
J^\m_\b = \stw (\s^\n \sb^\m \psi)_\b \del_\n \Ab -
\stw i \cWb' (\s^\m \psib)_{\b}, 
\ee    
from which the supercharge
\be 
Q_\a = \int d^3 x J^0_\a 
\ee  
is obtained. Note that the form of
the supercurrent and the supercharge are not modified  by
$C^{\a\b}$. Quantizing the fermions using the equal time
anticommutation relation, 
\be 
\{ \psi_\a (t,\vx), \psib_\da (t,\vx') \} = \delta_{\a\da} \delta
(\vx-\vx') 
\ee 
and keeping carefully the
boundary terms, one obtains for the anticommutator of two supercharges
\be \label{QQ-alg-wz} 
\{Q_\a, Q_\b\} =  4 i (\vs)_{\a\b} \cdot  \int
d^3 x \vec{\nabla} \cWb(\Ab).  
\ee 
Here $\vs_{\a\b}$ is defined by
$\vs_{\a\b} := \vs_{\a \db} \e^{\db\b}$ and is symmetric. Explicitly
$\vs_{\a\b} = \{ - \t^3, i \bb1, \t^1 \}_{\a \b}$.

The right hand side above is the central charge to the unbroken
$\cN=1/2$ supersymmetry algebra. It is a surface term which is
normally zero. However  the expression is nonzero in the presence of a
domain wall.  The value of the central charge is proportional to the
difference between the vacuum expectation values of $\cW$ in the two
distinct vacua between which the domain wall lies. For a wall lying in
the $xy$-plane, we have 
\be 
\{Q_\a, Q_\b\} = 2 i (\t_1)_{\a\b} \S {\sf A},
\ee 
where  ${\sf A}$ is the area of the
wall and
\be
\S  : = 2 \cWb(z=\infty) - 2 \cWb(z= - \infty) 
\ee 
is the central charge per unit area. 
Hence we have shown that central extension of the $\cN =1/2$
supersymmetry is possible.  Note that the result \eq{QQ-alg-wz} takes
the same form as in  the undeformed case with $C^{\a\b} =0$.  However
since the equations of motion are modified, the domain wall
configuration as well as the values of $\Wb$ will be modified in
general. Our next task is to solve 
\be \label{eom-wz-C} 
\del_z^2 \Ab = \cWb' \cW'', \quad  
\del_z^2 A =  \cWb'' [ \cW' + \l \det C (\cWb')^2]
\ee 
for the domain wall.

In the undeformed case $C^{\a\b} =0$, the equations reduce to the
form 
\be 
\del_z^2 \Ab = \cWb' \cW'', \quad 
\del_z^2 A = \cW' \cWb''
\ee 
for a domain wall extending in the $z$-direction. These second
order equations follows from the first order ones 
\be \label{eom-pair}
\del_z \Ab = e^{i \b} \cW'(A), \quad \del_z A = e^{-i \b} \cWb'(\Ab),
\ee 
where $\b$ is a constant phase factor. Moreover for real 
\be \label{real} 
A= \Ab, 
\ee  
the domain wall satisfies a single first order equation
\footnote{In this case $e^{i \b} = \pm 1$. Moreover we can always
choose $e^{i \b} = 1$ by absorbing the sign into $z$.}  
\be \label{eom-single} 
\del_z A = \cW'(A).  
\ee 
For example, for the
superpotential \eq{cubic-W} with $\m =\mub, \l = \lamb$,  we have the
solution 
\be \label{A0} 
A = \frac{\m}{\sqrt{\l}} \tanh (\mu \sqrt{\l} (z-z_0) ) .  
\ee 
This domain wall interpolates between the two different
vacua $A = \pm \m/\sqrt{\l} $ and has a central charge 
\be \label{CC}
\S = \frac {8 \m^3}{3 \sqrt{\l}}.  
\ee 

For the non-anticommutative case, the equation of motion can no longer be
reduced to first order form  as in \eq{eom-pair}. Also obviously one
cannot impose the reality condition \eq{real} anymore. Thus one has to
solve for the second order equations \eq{eom-wz-C} directly.  We begin
with an  analysis of the vacuum configurations. The classical
potential energy of the theory is given by 
\bea 
V &=& - \Fb F - F\cW'
- \Fb \cWb' - \frac{1}{3} \l \det C \;F^3 \nn\\ 
&=& \cWb' [\cW' +
\frac{1}{3} \l \det C (\cWb')^2] \label{V-WZ}.  
\eea 
The  vacuum
configurations satisfy $\del V /\del A=0 =\del V /\del \Ab$ and one
has the possibilities: \footnote{Note that in the first reference of 
\cite{renorm1}, it  was
assumed the quantity $H(A,\Ab)$ appearing  in $V = \cWb' [ \cW' -
H(A,\Ab)]$   has a nontrivial dependence on $A$. This is not the case
for our $V$ in \eq{V-WZ}. }  
\bea 
{\rm  (i):}  && \cWb'=0, \quad \cW'=0, \\ 
{\rm (ii):} && \cW''=0, \quad \cWb'' =0, \\ 
{\rm (iii):} && \cW''=0,\quad \cW' + \l \det C (\cWb')^2 =0, 
\eea 
which implies 
\bea
{\rm  (i):}  && A = \pm \frac{\m}{\sqrt{\l}}, \quad  \Ab = \pm
\frac{\mub}{\sqrt{\lamb}}, \quad \mbox{and}\quad V=0 , \label{vev1}\\
{\rm (ii):} && A= \Ab =0,  \quad \mbox{and}\quad V = \mub^2( \m^2+
\frac{\l \mub^4 }{3} \det C ), \label{vev2}\\ 
{\rm (iii):} && A=0,
\quad \Ab^2 =  (\mub^2  \pm \sqrt{\frac{\mu^2}{- \l    \det C}}) /\; \lamb,  
\quad \mbox{and}\quad  
V = \mp \frac{2\m^2}{3} \sqrt{\frac{\mu^2}{- \l \det C}}.\label{vev3} 
\eea 
The case (iii) is possible only if $\det C <0$,  in which cases we can 
have new vacuum configuration with energies less than zero.

Despite their more complicated form,  we are able to solve for and write
down several  explicit solutions of \eq{eom-wz-C}. For  simplicity, we
will take $\m = \mub, \l = \lamb$ below.  
Also since $A$ and $\Ab$ can vary independently 
without any reality constraint, we may take $\Ab$ to be sitting at
the vev  while we allow $A$ to vary. This is possible for 
$\Ab = - \mu/\sqrt{\l}$ and with $A$ obeying
\be 
\del^2 A = 2 \mu \sqrt{\l} \frac{\del \cW}{\del A}.  
\ee 
This can be integrated to give 
\be \label{Wk} 
\del A = \sqrt{4\mu \sqrt{\l}}\sqrt{\cW+ k}
\ee 
with an integration constant $k$;
or equivalently 
\be \label{Wei-P1}
\int_\infty^A \frac{dA}{\sqrt{4A^3 - g_2 A -g_3}} 
= - \sqrt{\frac{\m \l^{3/2}}{3}}(z-z_0), 
\ee 
where 
\be \label{g23} 
g_2 = \frac{12 \m^2}{\l}, \quad g_3 = -\frac{12 k}{\l}.  
\ee 
For $k \neq 0$, the equation \eq{Wei-P1} has a
solution given in terms of the  Weierstrass elliptic function  
\be A=
\wp (- \sqrt{\m \l^{3/2}/3}\,(z-z_0); g_2,g_3 ) 
\ee 
if the parameters $g_2, g_3$ given in \eq{g23} are chosen such that (in
particular, a negative $k$ is needed), 
\be 
g_2 = 60 \sum_{n,m\neq 0}
\frac{1}{(2n\o_1+ 2m \o_3)^4}, \quad  g_3 = 140 \sum_{n,m\neq 0}
\frac{1}{(2n\o_1+ 2m \o_3)^6}, 
\ee 
for some half periods $\o_1,
\o_3$. This solution is singular at $z=z_0$ (and its images) and its
physical meaning is not directly clear.  For $k =0$, the equation
\eq{Wk} can be integrated directly and  a regular solution can be
written down   in terms of the Jacobi elliptic function ${\rm sn}$,
\be 
\label{jacobi} 
A = \sqrt{\frac{3 \m^2}{\l}}
{\rm sn}^2 (\frac{\m \sqrt{\l}}{3^{1/4}}z , i ).  
\ee
However this solution is not a domain wall and does not carry a central charge.


We are interested in the domain wall solution, particularly one which
carries  a nonvanishing central charge.  Due to the complexity of
\eq{eom-wz-C}, we are not able to construct such solutions
explicitly.  However their existence  is easy to demonstrate.  To
see this, let us try to solve \eq{eom-wz-C} perturbatively with the
expansion parameter $\varepsilon := \det C$. Let 
\bea 
&& A = A_0 +
\varepsilon A_1 + \cdots, \\ 
&& \Ab = \Ab_0 + \varepsilon \Ab_1 +
\cdots, 
\eea 
where $A_0 =\Ab_0$ is given by \eq{A0} and $\cdots$
denotes terms of  higher order in $\varepsilon$. Note that $A_0, \Ab_0
\to \pm \m/\sqrt{\l}$  as $z\to \pm \infty$. $A_1$ and $\Ab_1$ satisfy
\bea \label{DE-A1} 
&& \del^2 \Ab_1 = 4 \l \m^2 \tanh^2(\m \sqrt{\l} z)
\; \Ab_1  - 2 \l \m^2 {\rm sech}^2(\m \sqrt{\l} z) \; A_1, \\ 
&&
\del^2 A_1 = 4 \l \m^2 \tanh^2(\m \sqrt{\l} z) \; A_1  
- 2 \l \m^2 {\rm sech}^2(\m \sqrt{\l} z) \; \Ab_1 
- 2 \l^{3/2} \m^5  \tanh(\m \sqrt{\l} z) {\rm sech}^4(\m \sqrt{\l} z).  \nn 
\eea
One can easily show that there exists solutions such that 
\be 
A_1, \Ab_1 \sim  e^{- 2 \m \sqrt{\l} |z|}, \quad 
\mbox{as $|z|\to \infty$}, 
\ee 
and hence the
asymptotic values of $A,\Ab$ are not affected.  The analysis can be
easily extended to the  higher orders and we conclude that the  system
\eq{eom-wz-C} admits a domain wall solution which interpolate between
the two vacua $\pm \m/\sqrt{\l}$. This domain wall carries the same
central charge \eq{CC}.

In the undeformed case, domain wall satisfying \eq{eom-pair} is BPS
saturated and  preserve half of the $\cN=1$ supersymmetry. This can be
seen easily from the supersymmetry transformations
\eq{susy-transf-wz2} and \eq{susy-transf-wz5} 
of $\psi$ and $\psib$.  In fact for a domain wall
extending in the $xy$ directions, if \eq{eom-pair} is satisfied, then
two of the supersymmetries obeying 
\be 
\xib^{\da} = i e^{i \b} (\s^3)^{\da \a} \xi_\a, 
\ee 
are preserved, that is, a linear
combination of the $Q$ and $\Qb$ supersymmetries is
preserved. Because of the preserved supersymmetry, the 3-dimensional field theory 
on the domain wall has vanishing vacuum energy and thus the domain wall energy density 
is not renormalized. This property does not hold for the
the deformed case. In fact in this case, the set of equations
\eq{eom-pair} are no longer consistent with the equation of motion
\eq{eom-wz-C}.  Therefore the domain wall annihilates all the
supersymmetry. This is to be expected since all the $\Qb$'s are broken
in the $\cN=1/2$ supersymmetry and this is the reason why the equation of
motion  cannot be reduced to the first order form \eq{eom-pair}. 

\section{$\cN=1/2$ Gauge Theories: Konishi Anomaly and Central Charge }

Let us now discuss the case of $\cN =1/2$ gauge theory. In
\cite{seiberg}. it is shown that the vector superfield $V$ may be
modified with an additional $C$-dependent part such that the component
fields transform canonically under gauge transformation. In the
Wess-Zumino gauge, $V$ is given by 
\bea \label{V}
V(y,\th,\bar{\th})
&=& -\th\sigma^{\mu}\thb A_{\mu}(y) +i\th\th \thb
\lamb(y) -i \thb \thb\th^{\alpha} \left(\l_{\alpha}(y)
+\frac{1}{4}\epsilon_{\alpha\beta}C^{\beta\gamma}
\sigma^{\mu}_{\gamma\dot{\gamma}}\left\{ \lamb^{\dot{\gamma}},A_{\mu}
\right\}(y) \right) \nn\\
&&+\frac{1}{2} \th\th \thb \thb\left( D(y)
-i\partial_{\mu}A^{\mu}(y)\right).  
\eea 
$V$ transforms under gauge transformation as
\footnote{The exponential functions $e^V = \sum V^{*n}/n!$ is defined
with $*$-product.}  
\be 
e^V \to e^{- i \overline{\Lambda}}*e^V* e^{i \Lambda}.  
\ee 
The gauge transformation which preserves the gauge
\eq{V} is given by 
\bea \label{gauge-para} 
\Lambda(y,\th)&=& -\varphi(y),\nn\\ 
\overline{\Lambda}(\yb,\thb)&=&
-\varphi(\yb)-\frac{i}{2} \thb\thb C^{\mu\nu} 
\{\partial_{\mu}\varphi, A_{\nu}\}(\yb) \nn\\ 
&=& - \varphi( y) +2
i\th\sigma^\mu\thb \partial_\mu \varphi (y) - \th\th\thb\thb
\partial^2 \varphi(y) -\frac{i}{2} \thb\thb C^{\mu\nu} \{\partial_\mu
\varphi, A_\nu\}, 
\eea 
and the gauge transformation of the component fields are the standard ones: 
\bea \label{gauge-transf} 
&\d A_{\mu}=-2\del_{\mu}\varphi+i[\varphi,A_{\mu}],&  \quad
\d D = i[\varphi,D],  \nn\\ 
&\d \lamb=i [\varphi,\lamb], & \quad \d \l= i[\varphi,\l]. 
\eea 
Note that although the $C$-dependent part in
\eq{V} and \eq{gauge-para} does not take value in the Lie-algebra,
nevertheless the component fields transform correctly.
The chiral and antichiral field strength superfields are defined by
\be
W_\a =-\frac{1}{4} \Db\Db e^{-V} D_\a e^V, \quad 
\Wb_\da = \frac{1}{4} DD e^{V} \Db_\da e^{-V} 
\ee 
and transform as
\be \label{W-transf} W_\a \to e^{-i\L} * W_\a * e^{i\L}, \quad
\Wb_\da \to e^{-i\overline{\L}}* \Wb_\da *e^{i\overline{\L}}.  
\ee 
In terms of components, we have 
\bea
W_\a &=& W_\a(C=0) + \frac{1}{2}
C_{\mu\nu}\sigma^{\mu\nu~ \beta}_\a  \th_\b\lamb \lamb(y), \\ 
\Wb_\da &=&  \Wb_\da (C=0) - \thb\thb \left[ \frac{C^{\mu\nu}}{2}
\{F_{\mu\nu},\lamb_\da \} + C^{\mu\nu} \{ A_\nu, {\cal D}_\mu
\lamb_\da - \frac{i}{4}[A_\mu ,\lamb_\da]\} +\frac{i}{16}  |C|^2
\{\lamb \lamb, \lamb_\da\}\right],\nn 
\eea 
where 
\bea 
W_\a (C=0) &=&
-i \lambda_\alpha(y) + \left[ \delta_\alpha^\beta D (y)-i
\sigma^{\mu\nu~ \beta}_\a  F_{\mu\nu}(y)  \right] \th_\b  + \th\th
\s^\mu_{\a\da} {\cal D}_\mu \lamb^\da(y),  \\ 
\Wb^\da (C=0) &=& i
\lamb^\da + \left[\d^\da_\db D- i (\sb^{\m\n})^\da{}_\db F_{\m\n}
\right]\th^\db - \thb \thb (\sb^\m)^{\da\b}\cD_\m \l_\b.  
\eea

A general supersymmetric gauge theory will also has matter fields
represented by chiral and antichiral superfields. Chiral superfield
has the standard component expansion \eq{chiral-1}. For antichiral
field, it is most convenient to parametrize its component expansion in
such a way that the component fields transform with standard gauge
transformation. 
The precise form will generally depends on the representation  with
respect to the gauge group. For example,  let $S, T$ be chiral
superfield   in the fundamental and anti-fundamental representation of
the gauge group,
and $\Sb,\Tb$ be antichiral superfields  in the anti-fundamental and
fundamental representations correspondingly
\bea 
S \to e^{-i \L} *S, \quad && \Sb \to \Sb *e^{i \Lamb}, \label{S-transf}\\ 
T \to T * e^{i \L}, \quad && \Tb \to e^{-i \Lamb}* \Tb.\label{T-transf} 
\eea 
If we parametrize   the $\thb\thb$ component of  $\Sb, \Tb$ in the
following manner \cite{ito}, 
\be \label{Sbar} 
\Sb (\yb, \thb) = \Ab_\rs (\yb) + \sqrt{2}\, \thb \psib_\rs (\yb) + \thb\thb \left(
\Fb_\rs (\yb) + iC^{\mu\nu} \partial_{\mu} (\Ab_\rs A_{\nu}) (\yb) -
\frac{1}{4} C^{\mu\nu} \Ab_\rs A_{\mu} A_{\nu} (\yb) \right), 
\ee 
\be
\label{Tbar} \Tb (\yb, \thb) = \Ab_\rt  (\yb) + \sqrt{2}\, \thb
\psib_\rt  (\yb) + \thb\thb \left( \Fb_\rt  (\yb) + iC^{\mu\nu}
\partial_{\mu} (A_\n \Ab_\rt ) (\yb) - \frac{1}{4} C^{\mu\nu}  A_{\mu}
A_{\nu}\Ab_\rt  (\yb) \right),
\ee  
then  the component fields of
$S,T, \Sb, \Tb$  all have the standard gauge transformations 
\bea &&
\d f = i \varphi f, \quad \mbox{for} \quad f= A_\rs, \psi_\rs, F_\rs,
\Ab_\rt, \psib_\rt, \Fb_\rt \nn\\ 
&& \d f = -i f  \varphi,  \quad
\mbox{for} \quad f =  A_\rt, \psi_\rt, F_\rt, \Ab_\rs, \psib_\rs,
\Fb_\rs.  
\eea 
As we shall see later, the form of the supersymmetry
transformation \eq{susy8} of $\l_\a$ imposes that the gauge group has
to be $U(N)$. The supersymmetry transformation \eq{susy8} may be
modified to adapt for the case of $SU(N)$ \cite{renorm2}, 
however the $\cN=1/2$
superfield formulation requires further investigation. For simplicity,
we will take the gauge group to be $U(N)$ in this paper.

\subsection{SQCD}

A theory of particular interest is the SQCD with $U(N)$ gauge group
and $N_\rf$ flavors, with each flavor consisting of a pair of chiral
superfields $\{ S_\ri, T_\ri \} $ in the fundamental and
anti-fundamental representations
of the gauge group.  The superpotential consists of the mass term  
\be
\cW_m = \sum_{\ri=1}^{N_\rf} m_\ri T_\ri S_\ri 
\ee
plus matter
self-interaction terms.  Without loss of generality, let us consider
the case of a single flavor $\{S,T\}$.   The $\cN=1/2$ SQCD is given
by the Lagrangian 
\bea \label{sqcd-star-Lagrangian} 
\cL &=&
\frac{1}{16 kg^2} \left( \int d^2 \th \; \tr W^\a* W_\a + \int d^2
\thb \; \tr \Wb_\da* \Wb^\da   \right) + \int d^4 \th \; (\Sb*e^V*S +
T*e^{-V}*\Tb) \nn\\ 
&& + \int d^2 \th \; m T*S + \int d^2 \thb\;  m \Sb* \Tb, 
\eea  
where $k$ is the normalization of the Lie algebra
generators: $\tr(T^a T^b) = k \d^{ab}$.   
In terms of the component fields, the Lagrangian reads (up to total
derivatives) 
\bea \label{L-sqcd} 
\cL =&&   \frac{1}{16 kg^2} \tr\left(
-4 i \lamb \sb^\m \cD_\m \l - F^{\m\n}F_{\m\n} + \frac{i}{2}
F^{\m\n}F^{\r\s}\e_{\m\n\r\s} + 2 D^2  -2i C^{\m\n} F_{\m\n}
\lamb\lamb + \frac{|C|^2}{2} (\lamb\lamb)^2 \right)\nn\\ 
&&+ \Fb_\rs
F_\rs  - i \psib_\rs  \sb^\m \cD_\m \psi_\rs   - \cD_\m \Ab_\rs
\cD^\m A_\rs   + \frac{1}{2} \Ab_\rs  D A_\rs   +
\frac{i}{\sqrt{2}}(\Ab_\rs  \l \psi_\rs  - \psib_\rs  \lamb A_\rs )
\nn\\ 
&& \quad+ \frac{i}{2} C^{\mu\nu} \Ab_\rs  F_{\mu\nu} F_\rs
-\frac{\sqrt{2}}{2}C^{\a \b}\s^{\mu}_{\a \da} \cD_{\mu}\Ab_\rs  \;
\lamb^{\da} \psi_{\rs\b} - \frac{|C|^2}{16} \Ab_\rs  \lamb \lamb F_\rs
\nn\\
 &&+ F_\rt  \Fb_\rt    - i \psi_\rt  \s^\m \cD_\m \psib_\rt   -
\cD_\m A_\rt  \cD^\m \Ab_\rt   - \frac{1}{2} A_\rt  D \Ab_\rt   -
\frac{i}{\sqrt{2}}(\psi_\rt  \l \Ab_\rt  - A_\rt  \lamb\, \psib_\rt )
\nn\\ 
&&\quad + \frac{i}{2} C^{\mu\nu} F_\rt  F_{\mu\nu} \Ab_\rt
-\frac{\sqrt{2}}{2}C^{\a \b}\s^{\mu}_{\a \da} \psi_{\rt\b} \lamb^\da
\cD_{\mu}\Ab_\rt   - \frac{|C|^2}{16} F_\rt  \lamb \lamb \Ab_\rt
\nn\\ 
&&+ m \left [F_\rt  A_\rs  + A_\rt  F_\rs     - \psi_\rt
\psi_\rs  - \psib_\rs  \psib_\rt  + \left(\Fb_\rs  + iC^{\mu\nu}
\partial_{\mu} (\Ab_\rs  A_{\nu}) -  \frac{1}{4} C^{\mu\nu} \Ab_\rs
A_{\mu} A_{\nu} \right)\Ab_\rt  \right.\nn\\ 
&& \qquad + \left.
\Ab_\rs \left( \Fb_\rt  + iC^{\mu\nu} \partial_{\mu} ( A_{\nu} \Ab_\rt
) -  \frac{1}{4} C^{\mu\nu}  A_{\mu} A_{\nu} \Ab_\rt \right)   \right],
\eea 
where  
\bea  
&& \cD_\m \l = \del_\m \l+ \frac{i}{2}[A_\m, \l], \qquad  
F_{\m\n} = \del_\m A_\n -\del_\n A_\m +  \frac{i}{2} [A_\m,A_\n], 
\label{cov1}\\ 
&& \cD_\m \psi_\rs  = \del_\m \psi_\rs + \frac{i}{2}A_\mu \psi_\rs \,, 
\quad   \cD_\m A_\rs  = \del_\m A_\rs + \frac{i}{2}A_\mu A_\rs , \, 
\label{cov2}\\ 
&& \cD_\m \psi_\rt  = \del_\m \psi_\rt  - \frac{i}{2}\psi_\rt  A_\mu \, , 
\quad   \cD_\m A_\rt  = \del_\m A_\rt  - \frac{i}{2}A_\rt  A_\mu\, .
\label{cov3} 
\eea
Note that since the second term of \eq{sqcd-star-Lagrangian}
transforms as  $\tr \Wb_\da* \Wb^\da \to  \tr (e^{-i
\overline{\L}}*\Wb_\da* \Wb^\da  *e^{i \overline{\L}})$ and the
$*$-product of $\overline{\L}$ with the rest cannot be ignored, the
gauge invariance of $\cL$ in the superfield form
\eq{sqcd-star-Lagrangian} is not apparent.
Nevertheless  this term is gauge invariant up to a total derivative 
as it is clear from the component expression.

The $\cN=1/2$ supersymmetry transformation is given by  
\bea
\label{susy-transf-sqcd-1} 
\d_{\xi}A_{\mu}&=& -i\lamb\sb_{\mu}\xi, \label{susy1}\\ 
\d_{\xi}\l_\a&=& i\xi_{\a} D+ (\s^{\mu\nu}\xi)_\a\left(F_{\mu\nu}+
\frac{i}{2} C_{\mu\nu} \lamb\lamb\right) ,
\quad \d_{\xi}\lamb_{\da}=0, \label{susy2}\\
\d_{\xi}D&=& -\xi \sigma^{\mu}{\cal D}_{\mu}\lamb, \label{susy3} 
\eea
for the gauge multiplet,  and \cite{ito}
\bea\label{susy-transf-sqcd-2} 
\d_{\xi}A_\ri&=& \sqrt{2}\xi\psi_\ri\, ,\quad \d_{\xi}\Ab_\ri=0,  
\label{susy4}\\ 
\d_{\xi}\psi_{\ri \a}&=& \sqrt{2}\xi_\a F_\ri\,, 
\quad \d_{\xi}\psib_{\ri\da} =
-i\sqrt{2}{\cal D}_{\mu}\Ab_\ri (\xi\sigma^{\mu})_{\da}, \label{susy5}\\ 
\d_{\xi}F_\ri &=&0,\label{susy6}\\ 
\d_{\xi}\Fb_\rs  &=& -i\sqrt{2}{\cal D}_{\mu}\psib_\rs \sb^{\mu} \xi 
-i \Ab_\rs  \xi\l +C^{\mu\nu} 
\left\{
\partial_{\mu}  \left(\Ab_\rs  \xi \sigma_{\nu} \lamb \right)
-\frac{i}{2} \left( \Ab_\rs  \xi \sigma_{\nu} \lamb \right) A_{\mu}
\right\}, \label{susy7}\\ 
\d_{\xi}\Fb_\rt  &=& -i\sqrt{2}{\cal
D}_{\mu}\psib_\rt  \sb^{\mu} \xi +i \xi \l \Ab_\rt   
+C^{\mu\nu} \left\{ 
\partial_{\mu}  \left( \xi \sigma_{\nu} \lamb\, \Ab_\rt \right) 
+\frac{i}{2} A_\m \left( \xi \sigma_{\nu} \lamb \,\Ab_\rt \right) 
\right\}
\label{susy8}
\eea for the matters $(\ri = \rs,\rt)$.  Note that in the
transformation \eq{susy2} for $\l_\a$,  $\lamb \lamb$ is given by an
anticommutator of the Lie algebra, therefore the transformation for
$\l$ is well defined only for $U(N)$.

\subsection{Konishi Anomaly}
 
Let us first review the undeformed case.  Let $S$ be any of the chiral
superfields of the SQCD.  Many years ago, it was realized 
\cite{sibold,kon1,kon2}
that the kinetic term of the chiral matter superfield
\be 
\cK :=\Sb e^V S
\ee 
satisfies the anomalous equations
\footnote{This result applies for more general  chiral or nonchiral
supersymmetric gauge theory.
For a chiral matter  $S$ in representation $R$, one has 
\be
\frac{1}{4}\Db^2 \cK = \tr\left( \frac{\del \cW}{\del S} S  \right) +
\frac{T(R)}{64 \pi^2} \, \tr(W^\a W_\a).  
\ee 
The factor $1/64\pi^2$ is with respect to the  normalization 
of the gauge multiplet in the
Lagrangian \eq{sqcd-star-Lagrangian}, \eq{L-sqcd}.  For the ``particle
physics normalization''  $1/4 F^{\m\n}F_{\m\n}$, we have to replace
$1/64 \pi^2 \to k g^2/16\pi^2$.  } 
\bea  
\frac{1}{4}\Db^2 \cK =
\frac{\del \cW}{\del S} S   + \frac{ 1}{64 \pi^2} \, \tr(W^\a W_\a),
\label{konishi-h} \\ \frac{1}{ 4}D^2 \cK =  \Sb \frac{\del \cWb}{\del
\Sb}   + \frac{ 1}{64 \pi^2} \, \tr(\Wb_\da \Wb^\da). \label{konishi-ah} 
\eea 
The first piece on the right hand
side of \eq{konishi-h}, \eq{konishi-ah} is classical.  It follows from
the classical invariance of the partition function of the theory under
the  infinitesimal rescaling of $S$ or $\Sb$ (all other fields kept fixed)
\bea \label{rescale-nostar-1} 
S \to  S \e &&\quad \mbox{for \eq{konishi-h}}, \nn\\ 
\Sb \to \bar{\e} \Sb &&\quad \mbox{for \eq{konishi-ah}}, 
\eea 
where $\e$ (resp. $\bar{\e}$) is an arbitrary
chiral (resp. anti-chiral) superfield.  The second piece has its
origin in the UV infinities which plague the composite operator
$\cK$. It is referred to  as the  Konishi anomaly and is of quantum
origin. For example in \eq{konishi-h}, there are UV divergences
in the operator product $\Fb_\rs A_\rs = - m A_\rt A_\rs $,  which
appears in the $\th=0$ component of  $-\frac{1}{4} \Db^2 \cK$ (see
\eq{D2K-1} below).  The UV divergences can be regulated.  However
additional contributions are induced after the  regulators are removed.

The Konishi anomaly can be computed by the standard techniques, such as
point splitting method, Pauli-Villars regularization or calculating
the  anomalous variation of the functional measure.
In nonchiral theory like the SQCD we consider here,  the Konishi
anomaly can be most  readily seen by using  the Pauli-Villars
regularization method.  The Pauli-Villars regulator fields  consist of
a pair of chiral superfields $\cQ, \cR$ of mass $M$ in the fundamental
and anti-fundamental representations.
As a result,   the operator product $m A_\rt A_\rs $ is replaced by
\be 
m A_\rt A_\rs  \to m A_\rt A_\rs  - M A_\cR A_\rQ.  
\ee 
The
Lagrangian is given by eqn. \eq{sqcd-star-Lagrangian}  with $m$
replaced by $M$ (and with $C=0$ for the undeformed case). The
contractions of the  fermion fields  are given by the Feynman rules
in figure 1. They can be obtained by first changing to a basis which
diagonalizes the Lagrangian of the fermions.
\begin{figure}
\label{fig1}
\psfrag{A}{${\displaystyle  = \frac{m}{p^2 + m^2}}$}
\psfrag{B}{${\displaystyle  =  \frac{m}{p^2 + m^2}}$}
\psfrag{C}{${\displaystyle  =  \frac{\sb^\m p_\m}{p^2 + m^2}}$}
\psfrag{D}{$ {\displaystyle  = \frac{\s^\m p_\m}{p^2 + m^2}}$}
\psfrag{psiS}{$\psi_\rs$} 
\psfrag{psibS}{$\psib_\rs$}
\psfrag{psiT}{$\psi_\rt$} 
\psfrag{psibT}{$\psib_\rt$}
\begin{center} 
{\scalebox{1}{ \includegraphics{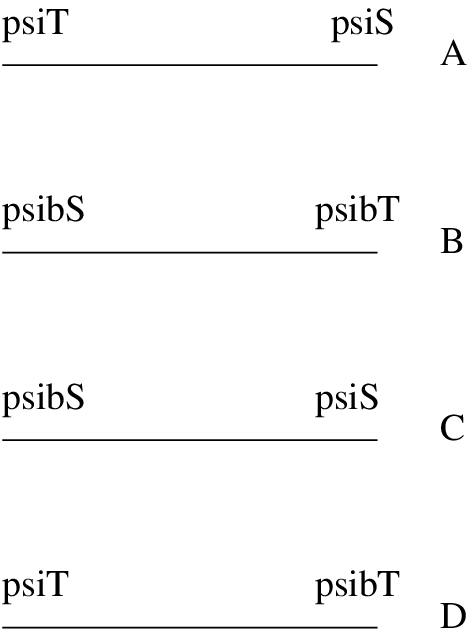}} }
\end{center}
\caption{Contractions for fermions in SQCD. }
\end{figure}
Integrating out the regulator fields in the gluino background, one can
show that an anomalous contribution arises from the triangular diagram
shown in figure 2. In the limit $M\to \infty$, we have 
\be 
\lim_{M\to \infty} \Fb_\rQ A_\rQ = \frac{1}{64 \pi^2} \tr (\l \l).  
\ee 
Similarly one can consider the other components of the superfield
$\frac{1}{4}\Db^2 \cK$  
and  find that $\tr (\l \l)$ is completed  to
$\tr(W^\a W_\a)$. Hence the result \eq{konishi-h} is obtained.
One can establish \eq{konishi-ah} in the same manner.

\begin{figure}
\label{fig2}
\psfrag{M}{$M$} 
\psfrag{QA}{$A_\rQ$} 
\psfrag{QAb}{$\Ab_\rQ$}
\psfrag{RA}{$A_\rR$} 
\psfrag{RAb}{$\Ab_\rR$} 
\psfrag{QF}{$F_\rQ$}
\psfrag{QFb}{$\Fb_\rQ$} 
\psfrag{RF}{$F_\rR$} 
\psfrag{RFb}{$\Fb_\rR$}
\psfrag{Qpsi}{$\psi_\rQ$}
\psfrag{Qpsib}{$\psib_\rQ$}
\psfrag{Rpsi}{$\psi_\rR$} 
\psfrag{Rpsib}{$\psib_\rR$} 
\psfrag{l}{$\l$}
\begin{center}
{\scalebox{0.8}{ \includegraphics{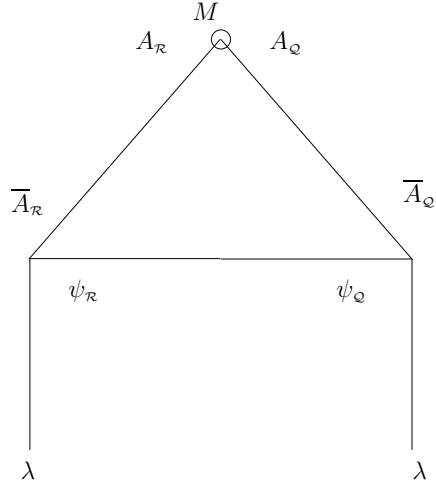}} }
\end{center}
\caption{Diagram contributing to the lowest component of the Konishi
anomaly in the undeformed case.}
\end{figure}

In the $\cN =1/2$ case, one considers the $*$-gauge invariant operator
\be  
\cK := \Sb * e^V * S.  
\ee    
From the classical invariance of the
partition function of the theory under the variation 
\be 
S \to S* \e,\quad \Sb \to \bar{\e}* \Sb, 
\ee
one obtains immediately 
\bea 
\label{konishi-nac-classical} 
&&\frac{1}{ 4}\Db^2 \cK =  \frac{\del \cW}{\del S} * S, \\ 
&& \frac{1}{4}D^2 \cK = \Sb * \frac{\del \cWb}{\del \Sb}  .  
\eea 
Next we compute the  quantum Konishi anomaly. 
Let us first consider $\frac{1}{4}\Db^2 \cK$.  To do
this, we need to first write down the component expansion of
$\frac{1}{4}\Db^2 \cK$ and then determine which composite  operators
get anomalous contribution.  The explicit expansion of  $\cK$ in
components is given in the appendix. It follows that 
\be \label{D2K-0}
\frac{1}{4}\Db^2 \cK = \frac{1}{4}\Db^2 \cK_0 + \frac{1}{4}\Db^2 \cE,
\ee 
where  
\bea  
\frac{1}{4}\Db^2 \cK_0 
&=&  \th^2 \left[ \Fb_\rs F_\rs  
+ (\cD_\m \cD^\m\Ab_\rs  - \frac{i}{\stw}\psib_\rs \lamb +
\frac{1}{2} \Ab_\rs D) A_\rs + i \cD_\m \psib_\rs \sb^\m \psi_\rs +
\frac{i}{\stw}\Ab_\rs \l \psi_\rs \right]\nn\\ 
&& + \th^\a \left[ \stw \Fb_\rs \psi_{_\rs\a} + 
\stw i  (\s^\m \cD_\m \psib_\rs)_\a A_\rs - i
\Ab_\rs \l_\a A_\rs \right] + \Fb_\rs A_\rs \label{D2K-1}\\
\frac{1}{4}\Db^2 \cE &=& \th^2 \cE_{22} + \th^\b (\cE_{12})_\b +
\cE_{02}. \label{D2K-2} 
\eea 
Here $\cE$ is given by \eq{K3} and 
$\cE_{22}, \cE_{12}, \cE_{02}$ by \eq{K4}. 
And  
\bea 
&& F_\rs = - m \Ab_\rt, \nn\\ 
&& \Fb_\rs = -m A_\rt -  \Ab_\rs( \frac{i}{2}C^{\m\n} F_{\m\n}  -
\frac{|C|^2}{16} \lamb\lamb).  
\eea
are to be substituted.
We remark
that one can also apply the equation of motion  to each of the
components of the $\th$ expansion, \eq{D2K-1} and  \eq{D2K-2}, and
establishes  the result \eq{konishi-nac-classical} explicitly. It is
quite nice that  the non-anticommutative $*$-product of the right hand
side of  \eq{konishi-nac-classical} is reproduced precisely.

Now let us turn to the computation of the Konishi anomaly.  It is
natural to guess 
that the term $\tr(W^\a W_\a)$ should be completed to
$\tr(W^\a * W_\a)$. We claim that this is indeed the case.  The
component expansion of $\tr(W^\a * W_\a)$ can be easily written down,
\bea \label{W-star-W} 
\tr(W^\a * W_\a) &=&   -\tr (\l\l) -
\frac{|C|^2}{4} \tr (\s^\m \cD_\m \lamb) (\s^\n \cD_\n \lamb) \nn\\ 
&&
-2 i \tr  \left[\l^\a(\d_\a^\b D - i \s^{\m\n}_\a{}^\b F_{\m\n}) -
\l_\a C^{\a\b}\lamb \lamb \right] \th_\b \nn\\ 
&& + \left[ \tr \left(
-2 i \lamb \sb^\m \cD_\m \l - \frac{1}{2} F_{\m\n} F^{\m\n} + D^2 +
\frac{i}{4}F^{\m\n}F^{\r\s} \e_{\m\n\r\s}\right) \right. \nn\\ 
&&
\left. \qquad - i C^{\m\n} \tr(F_{\m\n} \lamb \lamb) 
+ \frac{|C|^2}{4} \tr (\lamb\lamb)^2 \right]\th^2  .  
\eea  
Our task is to show that the
additional $C$-dependent terms are precisely generated in \eq{D2K-0}.
\begin{figure}
\label{fig3}
\psfrag{M}{$M$} 
\psfrag{QA}{$A_\rQ$} 
\psfrag{QAb}{$\Ab_\rQ$}
\psfrag{RA}{$A_\rR$} 
\psfrag{RAb}{$\Ab_\rR$} 
\psfrag{QF}{$F_\rQ$}
\psfrag{QFb}{$\Fb_\rQ$} 
\psfrag{RF}{$F_\rR$} 
\psfrag{RFb}{$\Fb_\rR$}
\psfrag{Qpsi}{$\psi_\rQ$} 
\psfrag{Qpsib}{$\psib_\rQ$}
\psfrag{Rpsi}{$\psi_\rR$} 
\psfrag{Rpsib}{$\psib_\rR$}
\psfrag{lamb}{$\lamb$} 
\psfrag{L}{$i \l_\a C^{a\b} \th_\b$}
\psfrag{Dlamb1}{$\frac{1}{\stw} C^{\c\d} \s^\m_{\c\dc} \cD_\m \lamb^\dc$} 
\psfrag{Dlamb2}{$\frac{1}{\stw} C^{\a\b} \s^\m_{\a\da} \cD_\m \lamb^\da$}
\begin{center}
{\scalebox{0.8}{ \includegraphics{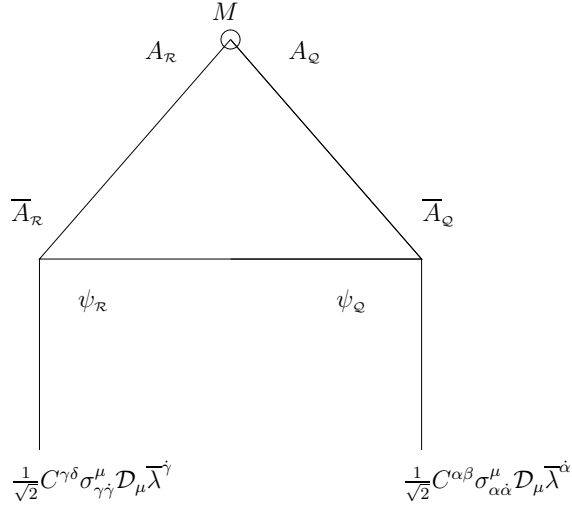}} }
\end{center}
\caption{Additional contribution   to the  Konishi anomaly at order
$\th^0$. }
\end{figure}

To see this, we have to examine carefully and determine which
operators pick up additional anomalous contribution.  At level
$\th^0$, we find that only the operator $m A_\rt A_\rs$ picks up an
anomalous contribution. The contributing diagram is shown  in figure
3.  At level $\th$, we list  in table 1 the  operators (of the regulator fields)
which contribute. 
They all contribute to generate the
operator 
\be 
c \times \frac{1}{64 \pi^2} i \l_\a C^{\a\b} \th_\b \lamb
\lamb 
\ee 
with coefficient $c$.  In total, we obtain $c=2$, which is
precisely what is needed in \eq{W-star-W}.  Finally, at level $\th^2$,
we list in table 2 the  operators (of the
regulator fields) which contribute
\footnote{Note that the operator $\Fb_\rs F_\rs +
(\frac{i}{2} C^{\m\n} \tr(F_{\m\n} \lamb \lamb) -\frac{|C|^2}{16}  \tr
(\lamb\lamb)^2) F_\rs = -m^2 A_\rt \Ab_\rt$.}.  They contribute to
generate the operator 
\be 
d_1 \times \frac{1}{64 \pi^2}
\left(\frac{i}{2} C^{\m\n} \tr(F_{\m\n} \lamb \lamb) -
\frac{|C|^2}{16}  \tr (\lamb\lamb)^2\right)   + d_2  \times
\frac{1}{64 \pi^2} \frac{|C|^2}{48} \tr (\lamb\lamb)^2 
\ee 
with
coefficients $d_1, d_2$.  Adding up their contributions, we get  
\be
\frac{1}{64 \pi^2} \left(  -i C^{\m\n} \tr(F_{\m\n} \lamb \lamb)
+\frac{|C|^2}{4}  \tr (\lamb\lamb)^2\right)  
\ee 
which is precisely
what is needed in \eq{W-star-W}. Therefore we obtain the result 
\be \label{konishi-nac-qm1} 
\frac{1}{4}\Db^2 \cK =  \frac{\del \cW}{\del S} * S   
+ \frac{1}{64 \pi^2} \tr(W^\a * W_\a) .  
\ee 

\begin{table}
\begin{tabular}{|c|c|c|}
\hline {}\hspace{3cm} operator \hspace{3cm}{} & contributing diagram &
coefficient $c$ \\ 
\hline&&\\ 
$ i M \th_\c C^{\c \a} \Ab_\rQ \l_\a \Ab_\rR$ & \mbox{figure 4a}&-1\\ 
$-\stw M A_\rR \th \psi_\rQ$ & \mbox{figure 4b}  & 2/3\\ 
$i \psib_\rQ \lamb \th_\c C^{\c \a}\psi_{\rQ \a}$ & \mbox{figure 4c}
  & 1
\\ $\stw \th_\c C^{\c \a} \Box \Ab_\rQ \psi_{\rQ\a}$ & \mbox{figure
  4d} & 2/3 
\\ $- i M \stw  \th_\c C^{\c \a} \e_{\a\k} (\sb^\m)^{\da \k} \del_\m
\psib_{\rQ\da} \Ab_\rR$ & \mbox{figure 4e}  & 2/3
\\ \hline
\end{tabular}
\caption{Operators contributing to Konishi anomaly at order
  $\th$. Deformed case. }
\end{table}
\begin{table}
\begin{tabular}{|c|c|c|c|}
\hline {}\hspace{3cm} operator \hspace{3cm}{} & contributing diagram &
\hspace{0.3cm} $d_1$ \hspace{0.3cm} &\hspace{0.3cm}$d_2$
\hspace{0.3cm}\\ 
\hline&&&\\ 
$M^2 A_\rR \Ab_\rR$ & \mbox{figure 5a}  & 0 & 1\\ 
$M^2 A_\rR \Ab_\rR$ & \mbox{figure 5b}  &  -1/3 & 0\\ 
$ -\frac{i}{\stw}\psib_\rQ \lamb A_\rQ$ & \mbox{figure 5c}  & -1 &0\\ 
$ -\frac{i}{\stw} \psib_\rQ \lamb A_\rQ$ & \mbox{figure 5d}& 0 &4 \\ 
$i \del_\m \psib_\rQ \sb^\m \psi_\rQ$ & \mbox{figure 5e}  &  0 &-3\\ 
$i \del_\m \psib_\rQ \sb^\m \psi_\rQ$ & \mbox{figure 5f}  & -2/3 &0\\ 
$ -\frac{1}{\stw} C^{\b\c}\s^\m_{\c\dc} \del_\m\Ab_\rQ \,\lamb^\dc
\psi_{\rQ\b}$  & \mbox{figure 5g}  &  0 &4\\ 
\hline
\end{tabular}
\caption{Operators contributing to Konishi anomaly at order
  $\th^2$. Deformed case. }
\end{table}

We note that
the relation \eq{konishi-nac-qm1} is gauge invariant.  
This can be seen either by
checking the component form, or by noting that the gauge parameter
$\L$ in the gauge transformation \eq{W-transf} is independent of $\th$
and hence insensitive to non-anticommutativity.  
We also remark that for the undeformed case,
it has been argued that the Konishi anomaly
satisfies an Adler-Bardeen theorem and is not renormalized beyond
1-loop \cite{ab}. It will be interesting to check it for the deformed case.
For $\frac{1}{4}D^2
\cK$, we note that  the (anti-holomorphic)  Konishi anomaly cannot be
simply given by $\tr(\Wb*\Wb)$ as it is not gauge invariant. This is
obvious due to the $\th$-dependence of $\Lamb$ and  the form of 
the gauge transformation \eq{W-transf}. The gauge non-invariance of
$\tr(\Wb*\Wb)$ can also be seen explicitly in the component form. For
example   at order $\thb^2$, $\tr(\Wb*\Wb)$ has a $C$-dependent part,
\be  \label{Wb-Wb} 
\tr \left( - i C^{\m\n} \tr(F_{\m\n} \lamb \lamb) +
\frac{|C|^2}{4} \tr (\lamb\lamb)^2 \right) -2i C^{\m\n}\del_\m
\tr(\lamb\lamb A_\n) 
\ee 
which is not gauge invariant. A natural guess
is that the  anti-holomorphic Konishi  anomaly is given by the gauge
invariant extension of  $ \tr(\Wb_\da * \Wb^\da)$. This is supported
by the fact that, apart from the total derivative term, the operators
in  \eq{Wb-Wb} are indeed generated at one loop by  exactly the same
set of diagrams in table 2.  As in the case of noncommutative gauge
theory, it may be possible to obtain the required gauge invariant
extension with the help of  Wilson line \cite{wilson}. 
It is also possible that the Adler-Bardeen theorem for the anti-holomorphic  
Konishi anomaly does not hold anymore and the higher loops contribute. 
We leave the  investigation of these issues for a further study.

\begin{figure}
\label{fig4}
\psfrag{M}{$M$} 
\psfrag{stw}{$\sqrt{2}$} 
\psfrag{thpsi}{$\th \psi_\rQ$} 
\psfrag{ithCpsi}{$i\th_\c C^{\c\a} \psi_{\rQ \a}$}
\psfrag{thCpsi}{$i\th_\c C^{\c\a} \psi_{\rQ \a}$}
\psfrag{delAb}{$\frac{1}{\stw} C^{\b\c}\s^\m_{\c\dc}\del_\m \Ab_\rR$}
\psfrag{thCe}{$ \stw \th_\c C^{\c\a} \e_{\a\k}$}
\psfrag{delpsib}{$i\ds \psib_\rQ$} 
\psfrag{boxA}{$\Box \Ab_\rQ$}
\psfrag{l}{$\l$} 
\psfrag{QA}{$A_\rQ$} 
\psfrag{QAb}{$\Ab_\rQ$}
\psfrag{RA}{$A_\rR$} 
\psfrag{RAb}{$\Ab_\rR$} 
\psfrag{QF}{$F_\rQ$}
\psfrag{QFb}{$\Fb_\rQ$} 
\psfrag{RF}{$F_\rR$} 
\psfrag{RFb}{$\Fb_\rR$}
\psfrag{Qpsi}{$\psi_\rQ$} 
\psfrag{Qpsib}{$\psib_\rQ$}
\psfrag{Rpsi}{$\psi_\rR$} 
\psfrag{Rpsib}{$\psib_\rR$}
\psfrag{lamb}{$\lamb$} 
\psfrag{L}{$i \l_\a C^{a\b} \th_\b$}
\psfrag{add}{$+$}
\begin{center}
{\scalebox{0.7}{ \includegraphics{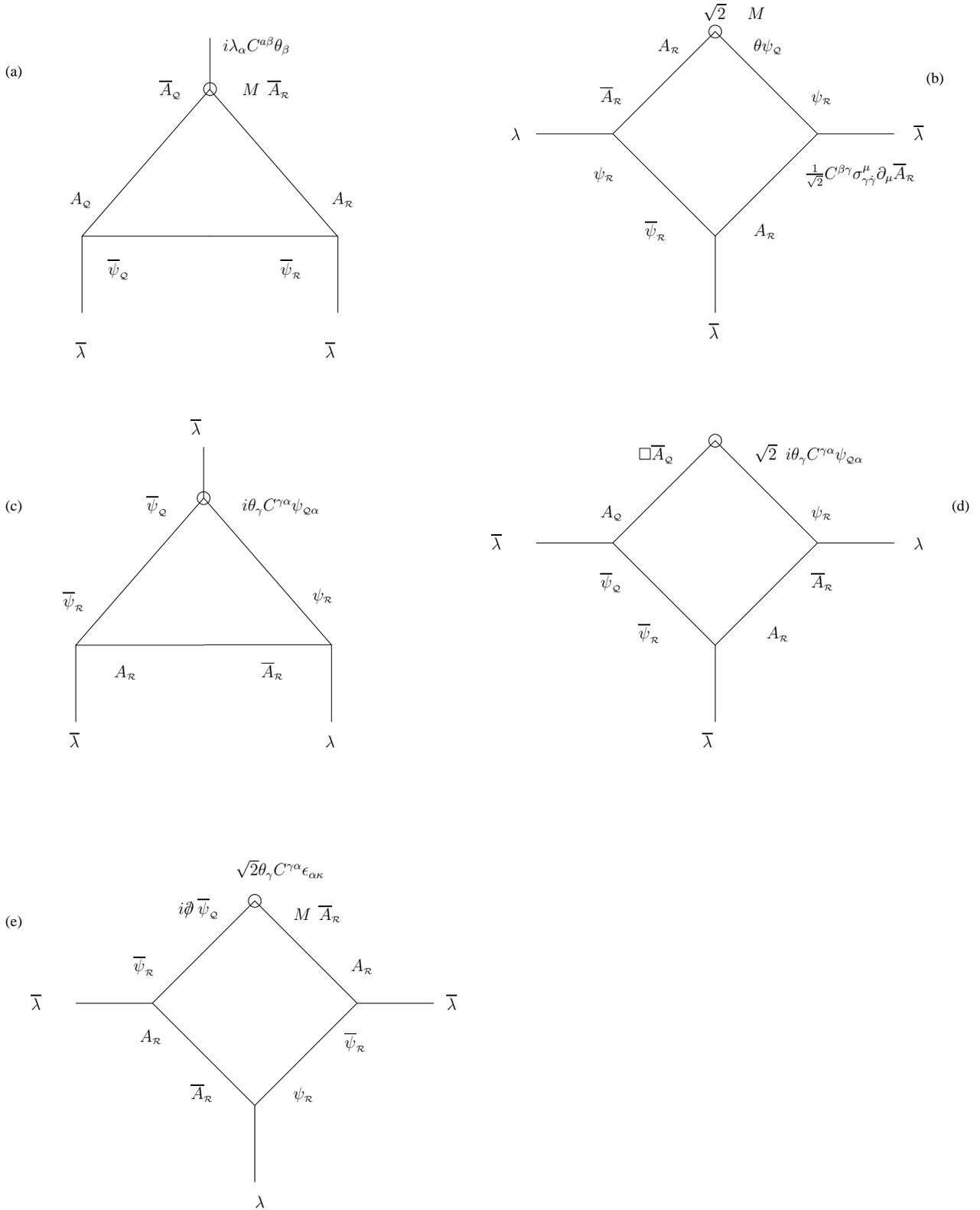}} }
\end{center}
\caption{Additional contributions to the  Konishi anomaly at  order
$\th^1$. }
\end{figure}

\begin{figure}
\label{fig5}
\psfrag{M}{$M$} 
\psfrag{M2}{$M^2$} 
\psfrag{stw}{$\sqrt{2}$}
\psfrag{Clamb1}{$-\frac{1}{\stw} C^{\a\b}\s^\m_{\a\da}\lamb^\da$}
\psfrag{Clamb2}{$- \frac{1}{\stw} C^{\c\d}\s^\m_{\c\dc}\lamb^\dc$}
\psfrag{dRAb}{$\del_\m \Ab_\rR$} 
\psfrag{dQAb}{$\del_\n \Ab_\rQ$}
\psfrag{CF}{$\frac{i}{2} C^{\m\n} F_{\m\n} - \frac{|C|^2}{16} \lamb\lamb$} 
\psfrag{MRAb}{$M \Ab_\rR$} 
\psfrag{delAb}{$\frac{1}{\stw}C^{\b\c}\s^\m_{\c\dc}\del_\m \Ab_\rR$} 
\psfrag{thCe}{$i \stw \th_\c C^{\c\a} \e_{\a\k}$} 
\psfrag{delpsib}{$i \ds \psib_\rQ$}
\psfrag{boxA}{$\Box \Ab_\rQ$} 
\psfrag{l}{$\l$} 
\psfrag{QA}{$A_\rQ$}
\psfrag{QAb}{$\Ab_\rQ$} 
\psfrag{RA}{$A_\rR$} 
\psfrag{RAb}{$\Ab_\rR$}
\psfrag{QF}{$F_\rQ$} 
\psfrag{QFb}{$\Fb_\rQ$} 
\psfrag{RF}{$F_\rR$}
\psfrag{RFb}{$\Fb_\rR$} 
\psfrag{Qpsi}{$\psi_\rQ$}
\psfrag{Qpsib}{$\psib_\rQ$} 
\psfrag{Rpsi}{$\psi_\rR$}
\psfrag{Rpsib}{$\psib_\rR$} 
\psfrag{lamb}{$\lamb$} 
\psfrag{L}{$i \l_\a C^{a\b} \th_\b$} 
\psfrag{add}{$+$}
\begin{center}
{\scalebox{0.8}{ \includegraphics{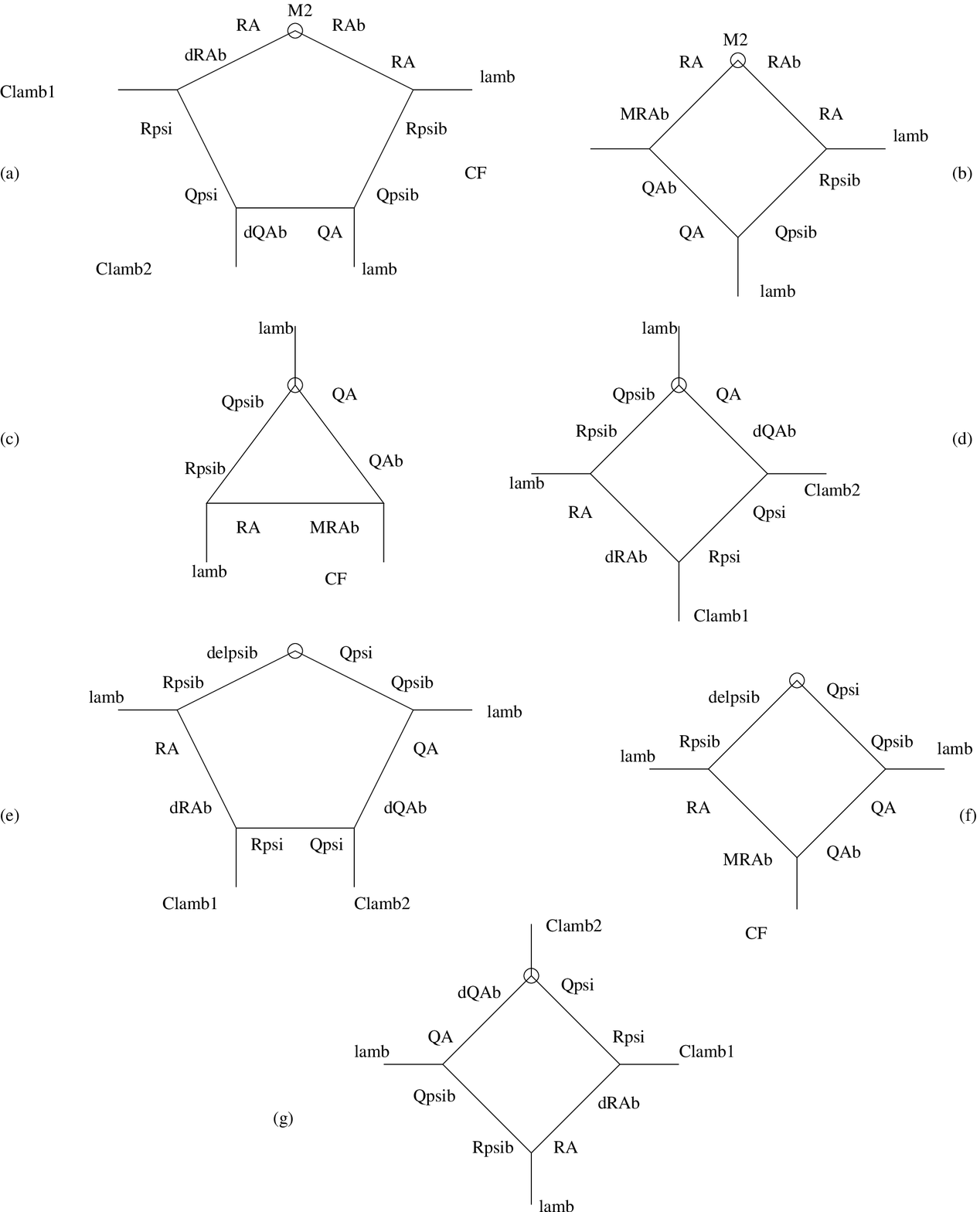}} }
\end{center}
\caption{Additional contributions  to the  Konishi anomaly  at  order
$\th^2$.}
\end{figure}

\subsection{Central Charge}
 
Let us now consider the central extension  in the SQCD
\eq{sqcd-star-Lagrangian}. To do this, we need to derive the form of
the supercharges, or the supercurrent.  Under the supersymmetry
transformations \eq{susy-transf-sqcd-1}, \eq{susy-transf-sqcd-2}, the
Lagrangian \eq{L-sqcd}  changed by a total derivative, from which one
can derive the supercurrent 
\bea \label{J-sqcd} 
J_\a^\m &=&
\frac{-i}{4 kg^2} ( \lamb \sb^\m \s^{\rho \n})_\a \left(F_{\rho \n} +
\frac{i}{2} C_{\rho \n} \lamb \lamb\right) \nn\\ 
&+& \sqrt{2}\cD_\n
\Ab_\rs  (\s^\n\sb^\m \psi_\rs )_\a  + \frac{1}{2} \Ab_\rs  (\s_\m
\lamb)_\a A_\rs   + \sqrt{2} (\s^\n\sb^\m \psi_\rt )_\a \cD_\n \Ab_\rt
+ \frac{1}{2} A_\rt  (\s_\m \lamb)_\a \Ab_\rt  \nn\\ 
&-& m \left( i
\sqrt{2} (\s^\m \psib_\rs )_\a \Ab_\rt  +i \sqrt{2} \Ab_\rs  (\s^\m
\psib_\rt )_\a + 2C^{\m\n} \Ab_\rs  (\s_\n \lamb)_\a \Ab_\rt  \right).
\eea 
Note that unlike the case of the Wess-Zumino model,  the current
is modified by $C^{\a\b}$. The modification is due to the additional
terms  in the supersymmetry transformations \eq{susy1}-\eq{susy8} that
are needed in order to keep the WZ gauge.  However like  the case of
the Wess-Zumino model, the form of the commutator   $\{Q_\a, Q_\b\}$
is not modified by $C$. We have 
\be \label{QQ-alg-sqcd} 
\{Q_\a, Q_\b\}
=  -4 i (\vs)_{\a\b} \cdot  \int d^3 x \vec{\nabla} (m \Ab_\rs \Ab_\rt ).  
\ee  
The above contribution to the central charge
is classical and came from  the superpotential $\cWb = m \Sb * \Tb$.
In addition to this classical contribution, there are additional
contributions  quantum mechanically. First there is a  contribution
from the usual  Konishi anomaly.  Indeed the operator $m \Ab_\rs
\Ab_\rt$ is the lowest component of  $-\frac{1}{4} D^2 \cK$. This
operator pick up a quantum contribution
\footnote{Note that here there is no analogous contribution as the one
in figure 3 because the counterpart of the coupling
$-\frac{1}{\stw}C^{\a \b}\s^{\mu}_{\a \da} \cD_{\mu}\Ab \; \lamb^{\da}
\psi_{\b}$ is absent in the Lagrangian \eq{L-sqcd}.}  from the diagram
in figure 6. We obtain   
\be 
m \Ab_\rs  \Ab_\rt  \to m\Ab_\rs  \Ab_\rt
-  \frac{1}{ 64 \pi^2} \tr(\lamb \lamb).  
\ee 
In addition to this
contribution which has a origin of Konishi anomaly, there is also a
contribution from the anomaly in the supercurrent. In the undeformed
case, this gives rise to  $\frac{N}{64 \pi^2} \tr(\lamb \lamb)$.  A
study of anomaly in the supercurrent multiplet in $\cN=1/2$ gauge
theory is in progress and  we expect the same contribution to the
central charge  as in the undeformed case \cite{progress}. Assuming
this is the case, we obtain the central extension for the $\cN=1/2$
SQCD, 
\be \label{CC-sqcd} 
\{Q_\a, Q_\b\} =  -4 i (\vs)_{\a\b} \cdot
\int d^3 x \vec{\nabla} \left(  m \Ab_\rs  \Ab_\rt  -
\frac{N-N_\rf}{64\pi^2}\, \lamb \lamb \right).  
\ee 
Note that this is
the same form as in the undeformed case.
Gluino condensate in $\cN =1/2$ gauge theory has been examined in 
in \cite{imaa} and it has been found that their values are unmodified by $C^{\a\b}$.
However as in the case of
the Wess-Zumino model, the value of the central charge may depends on
$C$ through the scalar profile.

\begin{figure}
\label{fig6}
\psfrag{M}{$M$} 
\psfrag{QA}{$A_\rQ$} 
\psfrag{QAb}{$\Ab_\rQ$}
\psfrag{RA}{$A_\rR$} 
\psfrag{RAb}{$\Ab_\rR$} 
\psfrag{QF}{$F_\rQ$}
\psfrag{QFb}{$\Fb_\rQ$} 
\psfrag{RF}{$F_\rR$} 
\psfrag{RFb}{$\Fb_\rR$}
\psfrag{Qpsi}{$\psi_\rQ$} 
\psfrag{Qpsib}{$\psib_\rQ$}
\psfrag{Rpsi}{$\psi_\rR$} 
\psfrag{Rpsib}{$\psib_\rR$}
\psfrag{lamb}{$\lamb$} 
\psfrag{Dlamb1}{$\frac{1}{\stw} C^{\c\d} \s^\m_{\c\dc} \cD_\m \lamb^\dc$} 
\psfrag{Dlamb2}{$\frac{1}{\stw} C^{\a\b} \s^\m_{\a\da} \cD_\m \lamb^\da$}
\begin{center} 
{\scalebox{0.8}{ \includegraphics{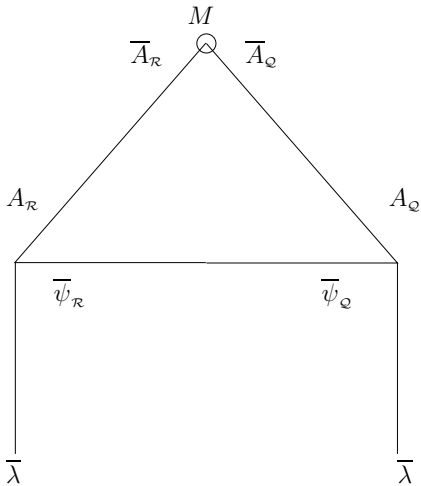}} }
\end{center}
\caption{Anomalous contribution to central charge. }
\end{figure}

\section{Discussion}

In this paper, we have shown that it is possible to centrally extend the
$\cN=1/2$ supersymmetry algebra and we determine the field theoretic 
form of the central extension in Wess-Zumino model and $\cN=1/2$ 
supersymmetric gauge theory. The domain wall we constructed satisfies
asymptotically $A =\Ab$  and reduces to the standard domain wall solution 
when $C\to 0$. It has a central charge 
independent of $C$. In principle it is possible to construct domain wall 
with more general asymptotic behaviour (i.e. tending to different vev's).
It is interesting to construct such more general domain walls.
As the $\cN=1/2$  supersymmetric gauge theory 
can be constructed as gauge theory on
D-brane, it is interesting to understand dynamical aspects such as
confinement, mass gap and chiral symmetry breaking, 
in terms of a D-brane construction \cite{d-qcd}.

We have also established the form of the holomorphic Konishi
anomaly. For the anti-holomorphic one, we show that the naive
extension has to be modified and we suggest that the
correct form is to be given by a gauge invariant completion of the term
$\tr \Wb* \Wb$. It will be interesting to perform a full analysis of this.
Konishi anomaly is related to the anomaly of the supercurrent
multiplet. It is also very interesting 
to determine the structure of the non-anticommutative
anomaly supermultiplet. 

Konishi anomaly has many physical applications. In this paper we
discuss its relation with the central charge of the $\cN=1/2$ supersymmetric
gauge theory. 
We expect that the more general form of the Konishi
anomaly \cite{cdsw} 
will shed light on a deformed version of the  Dijkgraff-Vafa theory \cite{dv}.

Given now the much nicer result for
the lower dimensional nonlinear sigma model \cite{abbp,agvm,ch},  
it will be interesting to determine the
condition for the vanishing of the one loop beta function and see how
the usual Ricci flatness condition is modified by non-anticommutativity.


\section*{Acknowledgements} 
CSC would like to thank the participants of the Bayrischzell Workshop  2005
for interesting discussions and comments, where the results of the paper 
were presented.
The authors would like to thank Ko Furuta and Valya Khoze 
for helpful discussions.
CSC acknowledges the support of EPSRC through an advanced fellowship.
TI acknowledges supports from the grants from JSPS (Kiban B and C) and the Chuo
University grant for special research. 
The authors wish to thank Koryu Kyokai for the grant of Japan-Taiwan
science collaboration, which helped us meet and collaborate.
 
\appendix \section{Components form of $\cK$}

In this appendix, we give the components form of $\cK = \Phib * e^V *\Phi$, 
where $\Phi$ is a chiral superfield \eq{chiral-1} and $\Phib$ is of
the form \eq{Sbar}. We have
\be \label{K1}
\cK = \cK_0 + \cE 
\ee
where the $C$-independent part $\cK_0$ is
\bea \label{K2}
\cK_0 &=& \th^2 \thb^2\left[ \Fb F +
\cD_\m  \cD^\m\Ab A + i \cD_\m \psib \sb^\m \psi + 
\frac{i}{\stw} (\Ab \l \psi - \psib \lamb A) 
+ \frac{1}{2} \Ab D A \right] \nn\\
&&+ \thb^2\left[\Fb A + \stw \Fb \th \psi + \stw i \th \s^\m \cD_\m \psib
  A - i \th\, \Ab \l A \right] \nn\\
&&+\th^2 \left[ \Ab F + \stw \thb \psib F - \stw i \cD_\m \Ab \,\thb \s^\m
  \psi + i \thb \,\Ab \lamb A\right]\nn\\
&& +\Ab A + \stw \Ab \th \psi + \stw \thb \psib A + \th \s^\m \thb
(-2i \cD_\m \Ab A + \psib \sb_\m \psi ),
\eea
and the $C$-dependent part $\cE$ has the expansion
\be \label{K3}
\cE = \th^2 \thb^2 \cE_{22} + \th^\b \thb^2  (\cE_{12})_\b + \thb^2 \cE_{02} +
\th_\a \thb^\da (\cE_{11})^\a{}_\da + 
\thb^\da (\cE_{01})_\da,
\ee
where $\cE_{mn}$ denotes the coefficient of $(\th)^m (\thb)^n$ in $\cE$:
\bea \label{K4}
\cE_{22} &=&  \frac{i}{2}C^{\m\n} \Ab F_{\m\n} F -
  \frac{|C|^2}{16}\Ab \lamb\lamb F - \frac{1}{\stw}
  C^{\b\c}\s^\m_{\c\dc} \cD_\m\Ab \,\lamb^\dc \psi_\b, \nn\\
(\cE_{12})_\b &= &  \stw \e_{\b\c}C^{\c\a}\left[ 
\big(\cD_\m\cD^\m \Ab + \frac{1}{2} \Ab D - \frac{i}{\stw}\psib \lamb \big)
  \psi_\a + \frac{i}{\stw} \Ab \l_\a F + i \e_{\a\k} (\sb^\m)^{\da \k}
  \cD_\m \psib_\da F\right] \nn\\
&&\qquad  + \cD_\m \Ab \s^\m_{\c\dc} \lamb^\dc C^{\c \k}(C_{\k\b} F -
  \e_{\k\b}A) 
 + \frac{i}{\stw} C^{\m\n} \Ab F_{\m\n} \psi_\b - 
\frac{|C|^2}{8    \stw} \Ab \lamb\lamb \psi_\b, \nn\\
\cE_{02} &=&
i C^{\a\c}\s^\m_{\c\da}\cD_\m\psib^\da \psi_\a + \frac{1}{\stw}
C^{\a\c}C_{\c\k}\e^{\k \b} \s^\m_{\b\da} \cD_\m \Ab \lamb^\da \psi_\a
+ \frac{i}{2} C^{\m\n} \Ab F_{\m\n} A \nn\\
&& \qquad  
- \frac{|C|^2}{4}\left(\cD_\m\cD^\m \Ab \,F + \frac{1}{4} \Ab \lamb\lamb A
-\frac{i}{\stw} \psib \lamb F + \frac{1}{2} \Ab D F\right) - \frac{i}{\stw}
\Ab \l_\a C^{\a\b} \psi_\b,  \nn\\
(\cE_{11})^\a{}_\da &=& -2 i  C^{\b \a} \s^\m_{\b \da} \cD_\m \Ab
  \,F  + \stw i \Ab \l_\da C^{\a\b}  \psi_\b, \nn\\
(\cE_{01})_\da &=& i \stw \cD_\m \Ab \psi_\a C^{\a\c}\s^\m_{\c\da} 
+\frac{i}{4} |C|^2 \Ab \lamb_\da F.
\eea
We have dropped the subscript $\rs$ for the fields here.
Here $C_{\a\b} := \e_{\a \a'} \e_{\b\b'} C^{\a'\b'}$, $C^{\m\n} :=
C^{\a\b} \e_{\b\c} (\s^{\m\n})_\a{}^\c$, $|C|^2 := C^{\m\n}C_{\m\n} =
4 \det C$. 
The Konishi current $T*e^{-V}*\Tb$ for chiral field $T$ in
anti-fundamental representation can be similarly written down.

\newpage

\end{document}